# Experimental artefacts in undrained triaxial testing

S. Ghabezloo & J. Sulem
*Université Paris-Est, UR Navier, ERMES, Ecole des Ponts ParisTech, Marne la Vallée, France*

ABSTRACT: For evaluation of the undrained thermo-poro-elastic properties of saturated porous materials in conventional triaxial cells, it is important to take into account the effect of the dead volume of the drainage system. The compressibility and the thermal expansion of the drainage system along with the dead volume of the fluid filling this system, influence the measured pore pressure and volumetric strain during undrained thermal or mechanical loading in a triaxial cell. A correction method is presented in this paper to correct these effects during an undrained isotropic compression test or an undrained heating test. A parametric study has demonstrated that the porosity and the drained compressibility of the tested material and the ratio of the volume of the drainage system to the one of the tested sample are the key parameters which influence the most the error induced on the measurements by the drainage system.

## 1 INTRODUCTION

The undrained condition is defined theoretically as a condition in which there is no change in the fluid mass of the porous material. For performing an undrained test in the laboratory, this condition cannot be achieved just by closing the valves of the drainage system as it is done classically in a conventional triaxial system (Figure (1)). In a triaxial cell, the tested sample is connected to the drainage system of the cell and also to the pore pressure transducer. As the drainage system has a non-zero volume filled with water, it experiences volume changes due to its compressibility and its thermal expansion. The variations of the volume of the drainage system and of the fluid filling the drainage system induce a fluid flow into or out of the sample to achieve pressure equilibrium between the sample and the drainage system. This fluid mass exchanged between the sample and the drainage system modifies the measured pore pressure and consequently the measured strains during the test. Wissa (1969) and Bishop (1976) were the first who studied this problem for a mechanical undrained loading and presented a method for correction of the measured pore pressure. Ghabezloo and Sulem (2009, 2010) presented an extension to the work of Bishop (1976) to correct the pore pressure and the volumetric strain measured during undrained heating and cooling tests, as well as undrained compression tests, by taking into account the compressibility and the thermal expansion of the drainage system, the inhomogeneous temperature distribution in the drainage system and also the compressibility and the thermal expansion of the fluid filling the drainage system. The correction method depends also on the porosity, the compressibility and the thermal expansion of the tested porous material.

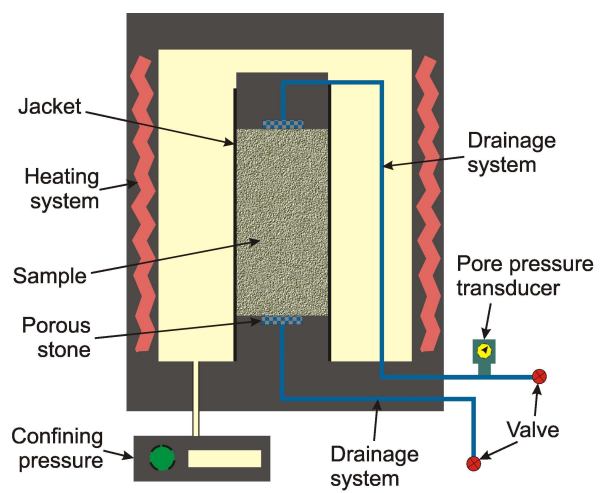

Figure 1. Schematic view of a conventional triaxial cell.

The proposed method was applied to the results of isotropic compression and undrained heating tests performed on Rothbach sandstone (Ghabezloo and Sulem 2009, 2010) and on a hardened cement paste (Ghabezloo et al., 2009). The proposed correction method is presented briefly in this paper.

## 2 POROELASTIC FRAMEWORK

We consider a fluid-saturated porous material with a porosity $\phi$. The variations of the total volume $V$ and of the pore volume $V_\phi$ are given as a function of variations of Terzaghi effective stress $\sigma_d$, pore pressure $p_f$ and the temperature $T$:

$$dV/V = -c_d d\sigma_d - c_s dp_f + \alpha_d dT \quad (1)$$

$$dV_\phi/V_\phi = -c_p d\sigma_d - c_\phi dp_f + \alpha_\phi dT \quad (2)$$

where $c_d$, $c_s$, $c_p$ and $c_\phi$ are four elastic compressibility coefficients, and $\alpha_d$ and $\alpha_\phi$ are two thermal expansion coefficients defined below:

$$c_d = -1/V \left(\partial V/\partial \sigma_d\right)_{p_f,T}, \quad c_p = -1/V_\phi \left(\partial V_\phi/\partial \sigma_d\right)_{p_f,T} \quad (3)$$

$$c_s = -1/V \left(\partial V/\partial p_f\right)_{\sigma_d,T}, \quad c_\phi = -1/V_\phi \left(\partial V_\phi/\partial p_f\right)_{\sigma_d,T} \quad (4)$$

$$\alpha_d = 1/V \left(\partial V/\partial T\right)_{p_f,\sigma_d}, \quad \alpha_\phi = 1/V_\phi \left(\partial V_\phi/\partial T\right)_{p_f,\sigma_d} \quad (5)$$

A detailed description of these parameters is presented in Ghabezloo et al. (2008,2009). In the undrained condition in which the mass of the fluid phase is constant ($dm_f = 0$), we can define four new parameters to describe the response of the porous material in undrained condition:

$$B = \left(\partial p_f/\partial \sigma\right)_{m_f,T}, \quad c_u = -1/V \left(\partial V/\partial \sigma\right)_{m_f,T} \quad (6)$$

$$\Lambda = \left(\partial p_f/\partial T\right)_{m_f,\sigma}, \quad \alpha_u = 1/V \left(\partial V/\partial T\right)_{m_f,\sigma} \quad (7)$$

The parameter $c_u$ is the undrained bulk compressibility, $B$ is the so-called Skempton coefficient, $\alpha_u$ is the undrained volumetric thermal expansion coefficient and $\Lambda$ is the thermal pressurization coefficient. Writing the fluid mass conservation under undrained condition ($dm_f = 0$) leads to the following expressions for the undrained parameters:

$$B = (c_d - c_s)/(c_d - c_s + \phi(c_f - c_\phi)) \quad (8)$$

$$\Lambda = \phi(\alpha_f - \alpha_\phi)/(c_d - c_s + \phi(c_f - c_\phi)) \quad (9)$$

$$c_u = c_d - B(c_d - c_s) \quad (10)$$

$$\alpha_u = \alpha_d + \Lambda(c_d - c_s) \quad (11)$$

where $c_f$ and $\alpha_f$ are respectively the pore-fluid compressibility and thermal expansion coefficient.

## 3 CORRECTION OF THE EFFECT OF DRAINAGE SYSTEM

In a triaxial cell the tested sample is connected to the drainage system and the undrained condition is achieved by closing the valves of this system (Figure (1)). Consequently, the condition $dm_f = 0$ is applied to the total volume of the fluid which fills the pore volume of the sample and also the drainage system ($m_f = V_\phi \rho_f + V_L \rho_{fL}$), where $\rho_{fL}$ is the density of the fluid in the drainage system and $V_L$ is its volume. As the sample and the drainage system may have different temperatures, temperature-dependent fluid densities are considered. The variation of volume of the drainage system is written as:

$$dV_L/V_L = c_L dp_f + \alpha_L dT_L - \kappa_L d\sigma \quad (12)$$

where $dT_L$ is the equivalent temperature change in the drainage system, $c_L$ and $\kappa_L$ are isothermal compressibilities and $\alpha_L$ is the thermal expansion coefficient of the drainage system defined as:

$$c_L = 1/V_L \left(\partial V_L/\partial p_f\right)_{T_L,\sigma} \quad (13)$$

$$\alpha_L = 1/V_L \left(\partial V_L/\partial T_L\right)_{p_f,\sigma} \quad (14)$$

$$\kappa_L = -1/V_L \left(\partial V_L/\partial \sigma\right)_{p_f,T_L} \quad (15)$$

In most triaxial devices, the drainage system can be separated into two parts: one situated inside the triaxial cell and the other one situated outside the cell. In the part inside the cell, the temperature change $dT$ is identical to the one of the sample; in the part situated outside the cell, the temperature change is smaller than $dT$ and varies along the drainage lines. We define an equivalent homogeneous temperature change $dT_L$ such that the volume change of the entire drainage system caused by $dT_L$ is equal to the volume change induced by the true non-homogeneous temperature field. The temperature ratio $\beta$ is thus defined as:

$$\beta = dT_L/dT \quad (16)$$

By writing the undrained condition $dm_f = 0$, using equation (12) the following expressions are obtained for the correction of the measured undrained thermo-poro-elastic parameters:

$$B^{cor} = \frac{B^{mes}}{1 + \frac{V_L \rho_{fL}}{V \rho_f (c_d - c_s)} \left[\kappa_L - B^{mes}(c_{fL} + c_L)\right]} \quad (17)$$

$$c_u^{cor} = c_d - \frac{c_d - c_u^{mes}}{1 + \frac{V_L \rho_{fL}}{V \rho_f (c_d - c_s)} \left[\kappa_L - \frac{c_d - c_u^{mes}}{c_d - c_s}(c_{fL} + c_L)\right]} \quad (18)$$

$$\Lambda^{cor} = \frac{\Lambda^{mes}}{1 + \frac{V_L \rho_{fL}}{V \rho_f \phi(\alpha_f - \alpha_\phi)} \left[\beta(\alpha_{fL} - \alpha_L) - \Lambda^{mes}(c_{fL} + c_L)\right]} \quad (19)$$

$$\alpha_u^{cor} = \alpha_d + \frac{\alpha_u^{mes} - \alpha_d}{1 + \frac{V_L \rho_{fL}}{V \rho_f \phi(\alpha_f - \alpha_\phi)} \left[\beta(\alpha_{fL} - \alpha_L) - (\alpha_u^{mes} - \alpha_d)\frac{c_{fL} + c_L}{c_d - c_s}\right]} \quad (20)$$

## 4 CALIBRATION OF THE CORRECTION PARAMETERS

The triaxial cell used in this study can sustain a confining pressure up to 60MPa. The axial and radial strains are measured directly on the sample with two axial transducers and four radial ones of LVDT type. The confining pressure is applied by a servo controlled high pressure generator. The pore pressure is applied by another servo-controlled pressure generator. The heating system consists of a heating belt around the cell which can apply a temperature change with a given rate and regulate the temperature, and a thermocouple which measures the temperature of the sample. More details about this triax-ial cell and a schematic view of the system are presented in Ghabezloo (2008).

The drainage system is composed of all the parts of the system which are connected to the pore volume of the sample and filled with the fluid, including pipes, pore pressure transducers, porous stones. The volume of fluid in the drainage system $V_L$, can be measured directly or evaluated by using the geometrical dimensions of the drainage system. For the triaxial cell used in the present study, the volume of the drainage system was measured directly using a pressure/volume controller equal to $V_L = 2300\,\text{mm}^3$.

The compressibility of the drainage and pressure measurement systems $c_L$ is evaluated by applying a fluid pressure and by measuring the corresponding volume change in the pressure/volume controller. A metallic sample is installed inside the cell to prevent the fluid to go out from the drainage system. Fluid mass conservation is written in the following equation which is used to calculate the compressibility $c_L$ of the drainage system:

$$dV_L/V_L = (c_L + c_{fL})dp_f \qquad (21)$$

where $dp_f$ and $dV_L$ are respectively the applied pore pressure and the volume change measured by the pressure/volume controller. For a single measurement, the volume change $dV_L$ accounts also for the compressibility of the pressure/volume controller and of the lines used to connect the pressure/volume controller to the main drainage system. This effect is corrected by performing a second measurement only on the pressure/volume controller and the connecting lines. The estimated value is $c_L = 0.117\,\text{GPa}^{-1}$.

The parameters $\beta$ and $\alpha_L$ are evaluated using the results of an undrained heating test performed using a metallic sample with the measurement of the fluid pressure change in the drainage system. For the metallic sample $\phi = 0$ and $c_d = c_s$ so that:

$$\Lambda^{mes} = \beta(\alpha_{fL} - \alpha_L)/(c_{fL} + c_L) \qquad (22)$$

The physical properties of water $\alpha_{fL}$ and $c_{fL}$ are known as functions of temperature and fluid pressure. As these variations are highly non-linear, the parameters $\beta$ and $\alpha_L$ cannot be evaluated directly but are back analysed from the calibration test results using equation (22). The parameters $\beta$ and $\alpha_L$ are back-calculated by minimizing the error between the measurements and the computed results using a least-square algorithm. $\beta$ is found equal to 0.46 and the thermal expansion coefficient of the drainage system $\alpha_L$ is found equal to $1.57 \times 10^{-4}\,(°C)^{-1}$.

The evaluation of the compressibility $\kappa_L$ which represents the effect of the confining pressure on the volume of the drainage system is performed using an analytical method. As can be seen in Figure (1), only a part of the drainage system which is the pipe connected to the top of the sample, is influenced by the confining pressure. The effect of the confining pressure on the variations of the volume of this pipe can be evaluated using the elastic solution of the radial displacement of a hollow cylinder and the following expression is obtained:

$$\kappa_L = 4\pi a^2 b^2 L(1-\nu^2)/((b^2 - a^2)V_L E) \qquad (23)$$

where $a$ and $b$ are the inner and the outer radius and $L$ is the length of the drainage pipe. $E$ and $\nu$ are the elastic parameters of the pipe material. For the dimensions of the triaxial system used in this study we obtain $\kappa_L = 1.6 \times 10^{-3}\,\text{GPa}^{-1}$ which is very small as compared to the compressibility $c_L = 0.117\,\text{GPa}^{-1}$. This is due to the fact that only a small part of the drainage system, less than 8% of its volume, is influenced by the confining pressure.

## 5 PARAMETRIC STUDY

Examples of the application of the proposed correction method on the results of the undrained isotropic compression test and undrained heating tests performed on a granular rock and a hardened cement paste are presented in Ghabezloo and Sulem (2009), Ghabezloo et al. (2009) and Ghabezloo and Sulem (2010). In this section, a parametric study on the error made on the measurement of different undrained thermo-poro-elastic parameters is presented. The error on a measured quantity $Q$ is evaluated as $(Q_{\text{measured}} - Q_{\text{real}})/Q_{\text{real}}$ and takes positive or negative values with indicates if the measurement overestimates or underestimates the considered quantity.

Among the different parameters appearing in equations (17) to (20), the porosity $\phi$ of the tested material, its drained compressibility $c_d$ and the ratio of the volume of the drainage system to the one of the tested sample, $V_L/V$ are the most influent parameters. For this parametric study the parameters of the drainage system are taken equal to the ones of the triaxial system used in this study. We take also $c_s = c_\phi = 0.02\,\text{GPa}^{-1}$ and $\alpha_\phi = 3 \times 10^{-4}\,(°C)^{-1}$ which are typical values. Figure (2) presents the error on the measurement of the Skempton coefficient as a function of the sample porosity, for three different values of drained compressibility and two different values of the ratio $V_L/V$. Three different values of the drained compressibility are considered, respectively equal to $0.03\,\text{GPa}^{-1}$, $0.1\,\text{GPa}^{-1}$ and $0.5\,\text{GPa}^{-1}$, which covers a range from a rock with a low compressibility to a relatively highly compressible rock. The porosity is varied from 0.05 to 0.35. The ratio $V_L/V$ is taken equal to 0.025 which corresponds to the conditions of the triaxial system used in this study. We analyze also the effect of a greater volume of the drainage system on the measurement errors by choosing another value twice bigger, equal to 0.05. We can see in Figure (2) that the error on the measurement of $B$ is always negative (the measurement underestimates the real value) and covers an important range between 2% and 50%. The measurement error is more significant for low-porosity rocks with low-compressibility. We can also see the significant effect of the volume of the drainage system on the measurement error. The measurement error for the undrained compressibility $c_u$ is presented in Figure (3) where we can observe that it is more important for low-porosity rocks and for greater volume of the drainage system. As opposite to what we observe for $B$, the error of the measurement of $c_u$ is more important when the tested material is more compressible.

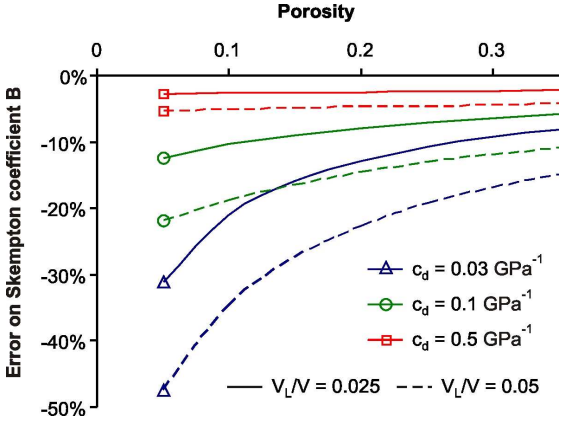

Figure 2. Error on Skempton coefficient $B$

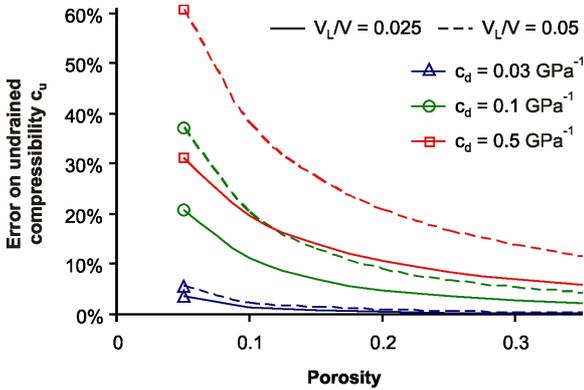

Figure 3. Error on the undrained compressibility $c_u$

Figures (4) and (5) show the errors corresponding to the measurements of the thermal pressurization coefficient $\Lambda$ and of the undrained thermal expansion coefficient $\alpha_u$ respectively. The error of the measurement for $\Lambda$ varies between -40% and +10%, which shows that the measured value may be smaller or greater than the real one. As for the isothermal undrained parameters, the error is more important for low-porosity materials and for a greater volume of the drainage system. The error for the undrained thermal expansion coefficient $\alpha_u$ varies between -6% and +4%, which is a narrower range, as compared to the other undrained parameters.

## 6  CONCLUSIONS

A simple method is presented for analysis of the error induced by the dead volume of the drainage system of a triaxial cell on the measurement of undrained thermo-poro-elastic parameters. A parametric study demonstrated that the porosity $\phi$ of the tested material, its drained compressibility and the ratio of the volume of the drainage system to the one of the tested sample, $V_L/V$ are the key parameters which influence the most the error induced on the measurements by the drainage system. It was also shown that the Skempton coefficient, the thermal pressurization coefficient and the undrained compressibility measurements are much more affected than the measurement of the undrained thermal expansion coefficient.

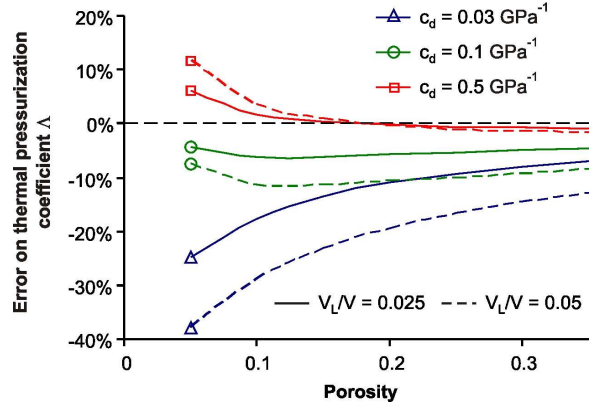

Figure 4. Error on the thermal pressurization coefficient $\Lambda$

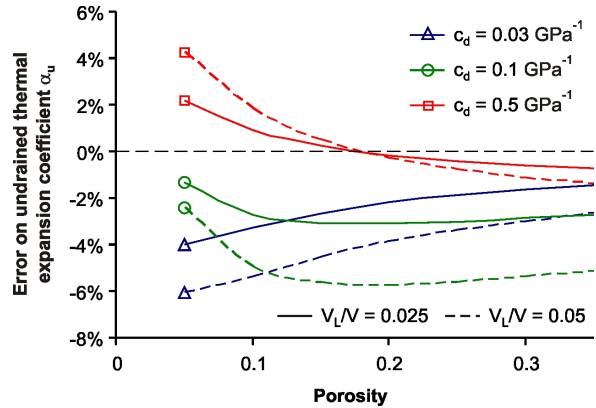

Figure 5. Error on the undrained thermal expansion coeff. $\alpha_u$